\documentclass[12pt]{iopart}

\usepackage{graphicx}

 \usepackage{subeqn}

\usepackage[T1]{fontenc}
\newcommand{\GN}{N}

\newcommand{\av}[1]{\ensuremath{\mbox{\guilsinglleft}#1\mbox{\guilsinglright}}}

\newcommand{\be}{\begin{equation}}
\newcommand{\ee}{\end{equation}}
\newcommand{\ba}{\begin{array}}
\newcommand{\ea}{\end{array}}
\newcommand{\bqa}{\begin{eqnarray}}
\newcommand{\eqa}{\end{eqnarray}}

\begin{document}

\title{High fidelity quantum gates of trapped ions in the presence of motional heating}
\author{Farhang Haddadfarshi and Florian Mintert}

\address{Department of Physics, Imperial College London, London SW7 2AZ, United
Kingdom}

\vspace{10pt}

\begin{abstract}
We describe entangling quantum gates for trapped ions mediated by a dissipative bus mode
and show that suitably designed, polychromatic control pulses decrease ion-phonon entanglement substantially while maintaining the mediated interaction.
In particular for multi-qubit gates this yields a significant improvement in gate performance. 
\end{abstract}
\newpage

\section{Introduction}

The realization of quantum gates is the fundamental building block in the exploitation of quantum mechanical systems for purposes like the execution of quantum algorithms, digital quantum simulations, quantum teleportation or precision sensing \cite{politi2009shor,digitalsimulation,riebe2004deterministic,barrett2004deterministic,giovannetti2004quantum}.
Some proof-of-principle experiments can be performed with mediocre quantum gates, but harnessing the true potential of quantum mechanical systems requires high-fidelity gates.
In the last years one could therefore witness dedicated efforts towards the increase of gate fidelities.
Single qubit gates for trapped ions have been implemented with infidelities reaching $10^{-6}$
\cite{PhysRevLett.113.220501}; but given the increased complexity, infidelities for entangling gates for two ions are substantially higher, in the range of $10^{-3}$ \cite{benhelm2008towards,PhysRevLett.117.060504}.

The potential of trapped ions for applications of quantum information theory has been demonstrated abundantly \cite{PhysRevLett.82.1835,haffner2005scalable,blatt2008entangled},
but the limited fidelities of entangling gates still pose an obstacle to fault tolerance.
Despite the fact that qubits can be encoded in highly stable hyperfine, dressed or clock states \cite{timoney2011quantum,PhysRevLett.110.263002} with very long coherence times,
entangling gates are limited by the comparatively short coherence times of the collective oscillation that is needed to mediate an interaction between the otherwise non-interacting qubit degrees of freedom.
Since the implementation of a gate necessarily implies some entanglement between the qubits and this bus mode, the decoherence of the latter unavoidably has a detrimental impact on the former.

Substantial progress towards the goal of making qubits independent of the bus  
mode decoherence was made in terms of the M\o lmer-S\o rensen (MS) gate \cite{molmer}.
It is independent of the state of the bus mode; that is, it is insensitive to any type of decoherence prior to the gate operation.
This feature has helped substantially towards the experimental implementation of two-qubit and multi-qubit gates \cite{sackett2000experimental,leibfried2003experimental,PhysRevLett.106.130506},
but sensitivity to bus mode decoherence during the gate operation is still a limiting factor.
As we show here, this sensitivity can be reduced substantially through suitably chosen polychromatic driving  even if decoherence of the bus mode can not be reduced.
As this improvement grows with the number of qubits involved in a gate, our approach promises to permit the high-fidelity realization of quantum algorithms or quantum simulations with imperfect hardware.

\section{Dissipative MS gate}

The MS gate is realised in terms of driving with slight detuning with respect to the red and blue sideband transitions respectively.
The corresponding Hamiltonian reads
\begin{equation}   
\mathcal{H}(t)=(\Upsilon(t)\ a +\Upsilon^{\ast}(t)\ a^{\dagger})\ \mathcal{S}_x\ ,\nonumber
\label{eq:HMS}
\end{equation}
in terms of
the creation and annihilation operators $a^{\dagger}$ and $a$ of phonons in the bus mode and
the collective operator $\mathcal{S}_x=\sum_j\sigma_x^{(j)}$ where $\sigma_x^{(j)}$ induces transitions between the two qubit states of the $j$-th ion.
The time-dependent amplitude $\Upsilon(t)$
 for a conventional MS gate
% typically
%{\bf{for a monochromatic gate}}
equals $\eta\Omega\exp(i\delta t)$ with the Lamb-Dicke parameter $\eta$, the Rabi frequency $\Omega$ and the detuning $\delta$ \cite{molmer},
but, in the following, we will consider more general time dependencies. The propagator induced by $\mathcal{H}(t)$ reads
\begin{equation}
\mathcal{U}_{\mathcal{K}}=\exp\left(-i\big((f(t)a+f^{\ast}(t)a^{\dagger})\mathcal{S}_x-g(t)\mathcal{S}_x^2\big)\right)
\label{eq:UMS}
\end{equation}
with $f(t)=\int_0^t dt^{\prime}\Upsilon(t^{\prime})$ and
$g(t)=\Im [\int_0^tdt^{\prime}\Upsilon(t^{\prime})f^\ast(t^\prime)]$.

At instances at which $f$ vanishes, this describes the dynamics induced by the spin-spin-type interaction $\mathcal{S}_x^2$ independently of the the motional state of the bus mode.
This feature permits implementation of multi-qubit quantum gates even with incoherent, thermal states, what has resulted in experimental implementations with high gate fidelities \cite{benhelm2008towards,kirchmair2009deterministic, PhysRevLett.109.020503}.
In the quest for gates that permit the realization of fault tolerant quantum computation, however, incoherent processes during the gate operation are a limiting factor \cite{PhysRevA.62.022311}. 
The qubits and motional degrees of freedom get correlated (as accounted for by the term $(f(t)a+f^{\ast}(t)a^{\dagger})\mathcal{S}_x$ in Eq.(\ref{eq:UMS}) so that dissipation of the ions' motion affects the coherence of the qubits.
Predominant effects are thermalisation and dephasing that can be modelled with a Master equation characterized by a
generator \cite{lindblad}
\begin{equation}
\mathcal{L}[\circ]=-i[\mathcal{H}(t),\circ]+\sum_{j=+,-,d}\gamma_j\mathcal{D}_{E_j}[\circ] \ .
\label{eq:lindblad}
\end{equation}
The first term $-i[\mathcal{H}(t),\circ]$ describes coherent dynamics and the dissipator is comprised of the terms $\mathcal{D}_{\hat{O}}[\circ]=\hat{O}\circ\hat{O}^{\dagger}-\frac{1}{2}\{\hat{O}^{\dagger}\hat{O},\circ\}$  with  operators
$E_-=a$ and $E_+=a^\dagger$ for thermalisation and  $E_d=a^\dagger a=\hat n$ for dephasing.

In the following we consider all dynamics in the time-dependent frame defined by $\mathcal{U}_{\mathcal{K}}(t)$.
In this frame, the ideal coherent dynamics, including the perfect, entangling gate is described by the identity ${\bf{1}}$, and the Master equation becomes a purely dissipative equation with a time-dependent  dissipator $\tilde{\mathcal{L}}[\circ]=\sum_j\gamma_j\mathcal{D}_{\tilde{E}_j(t)}[\circ]$ with
\begin{subequations}
\begin{eqnarray}
\tilde{E}_-&=&a-if^{\ast}(t)\mathcal{S}_x\ ,\\
\tilde{E}_+&=& a^\dagger+if(t)\mathcal{S}_x\ ,\\
\tilde{E}_d&=&\hat n+i(af(t)-a^\dagger f^\ast(t))\mathcal{S}_x+ |f(t)|^2\mathcal{S}_x^2\ .
\end{eqnarray}
\label{eq:linbladoperators}
\end{subequations}
The Lindblad operators $\tilde E_j$ directly affect both qubits and motional degrees of freedom, which is the formal manifestation of the fact that motional decoherence affects the qubits during gate operation.
In state of the art experiments the decay rates are sufficiently small such that $\gamma_jT\ll 1$,
where $T$ is the gate duration.
The propagator induced by the time-dependent generator $\tilde{\mathcal{L}}$ can thus be approximated perturbatively 
as $\tilde{\mathcal{V}}(t)\simeq {\bf{1}}+\int_{0}^{t}dt^\prime \tilde{\mathcal{L}}_{t^\prime}$,
and the infidelity ${\cal I}$ of the gate operation can be characterized (in leading order) by the magnitude of the map
\begin{equation}
\Xi(\circ)=
\tr_{{\cal B}}\int_{0}^{T}dt^\prime\ {\tilde{\cal L}}_{t^\prime}[\circ\otimes\varrho_\mathcal{B}]\ ,
\label{eq:map}
\end{equation}
where $\tr_{{\cal B}}$ denotes the trace over the bus mode with the density operator $\varrho_{\mathcal{B}}$.  
In practice, it is defined as ${\cal I}=||\zeta||^2=\tr\zeta^\dagger\zeta$
in terms of the norm of the matrix $\zeta$  \cite{PhysRevA.71.062310}, with elements
related to $\Xi$ via $\Xi(\circ)=\sum_{ij}\ \zeta_{ij}\ \xi_i\ \circ\ \xi_j^\dagger$.
The $N$-qubit operators utilized in this expression are defined as $\xi_0=\bf{1}$  and
\begin{equation}
\xi_i={N\choose i}^{-\frac{1}{2}}\sum_{1\le j_1<...<j_i\le N}\sigma_x^{(j_1)}...\sigma_x^{(j_i)} \nonumber
\label{eq:xi}
\end{equation}
for $i=1,2,3,4$.
 These operators describe simultaneous state changes on up to four ions,
which will result in a dependence of ${\cal I}$ on the ions number in terms of powers $N^i$ with $i\le 4$. 
The gate infidelity depends on the state of the bus mode via the properties $\langle\hat n\rangle$, $\langle a\rangle$, $\langle a^2\rangle$ as shown in more detail in the appendix.
Since, in practice, a control scheme is desirable that is independent of the state of the bus mode and the exact form of its dissipation, we will refrain from optimizing the gate fidelity for very specific situations, but rather target solutions that yield good performance largely independent of bus mode properties.
To this end, it is helpful to realize that $\zeta$ depends on the pulse $\Upsilon(t)$ via the functions $\av{f}$, $\av{|f|^2f}$, $\av{|f|^2}$ and $\av{|f|^4}$, where the symbol $\av{\cdot}=\int_{0}^{T}dt\ \cdot $ denotes time-integration over the gate duration $T$.
 Strictly speaking, the infidelity depends also on higher moments of $f$, but these dependencies appear only in higher orders than considered in Eq.(\ref{eq:map}) and, they can safely be neglected in the relevant regime of weak dissipation.

\section{Polychromatically driven quantum gates}

As deviations from perfect gate operations result from qubit-phonon entanglement as characterized by $f(t)$ in Eq.(\ref{eq:UMS}) one would expect to achieve high fidelities by minimizing the deviations of $f(t)$ from $0$.
It is always possible to choose $\Upsilon(t)$ such that $\av{f}$ and $\av{|f|^2f}$ vanish,
but since $\av{|f|^2}$ and $\av{|f|^4}$ are averages over non-negative functions,
they will always be non-negative for a non-vanishing pulse $\Upsilon(t)$.
We found that a large improvement is obtained if $\av{f}$ vanishes, but that the additional requirement that also $\av{|f|^2f}$ vanish yields hardly further improvement.
In fact, the second most relevant aspect is the minimization of $\av{|f|^2}$, and we observed better gate performances with a minimization of $\av{|f|^2}$ under the constraint $\av{f}=0$ rather than the two constrains $\av{f}=\av{|f|^2f}=0$.
Since the pulse $\Upsilon(t)$ obtained in this fashion also typically yields minimal values of $\av{|f|^4}$
this permits to identify driving patterns that achieve optimal gates, irrespective of the bus mode properties.

 For the identification of optimised pulses it is helpful to work with an explicit parametrization.
We strive for a Fourier series
\begin{equation}
\Upsilon_p=\sum_{j=1}^{m}c_j\omega\exp(ij\omega t)\ , \nonumber
\end{equation}
 where we define all amplitudes in multiples of the fundamental frequency $\omega$
in order to arrive at unitless, complex amplitudes $c_j$. The condition for ${\cal U}_{\cal K}(T)$ with $T=2\pi/\omega$ to be a maximally entangling gate $\exp(-i\frac{\pi}{8}{\cal S}_x^2)$ reads $g_T=\sum_j |c_j|^2/j=1/16$, and
the optimization to be solved reads
\be
\min_{\{c_j\}}\left(\left.\sum_{j=1}^{m}\frac{|c_j|^2}{j^2}\ \right|\ \sum_{j=1}^{m}\frac{|c_j|^2}{j}=\frac{1}{16}\ ,\ \sum_{j=1}^{m}\frac{c_j}{j}=0\right)\ \nonumber.
\label{optimization}
\ee
The optimal solution $c_j^{opt}$ \cite{PhysRevLett.113.010501}, that, in particular can be chosen real, reads
\be
c_j^{opt}=\frac{jb}{1-j\lambda}\ \mbox{ and }\ b=-\frac{1}{4}\left(\sum_{j=1}^{m}\frac{j}{(1-j\lambda)^2}\right)^{-\frac{1}{2}}\ ,
\label{eq:optimalpulse}
\ee
where $\lambda$ is the smallest root of the equation
\be
\sum_j^m(1-j\lambda)^{-1}=0\ .
\label{eq:lambda}
\ee
The function $\sum_j^m(1-j\lambda)^{-1}$ diverges for $\lambda\to 1/k$ with $k=1,...,m$ and the smallest root lies in the interval $[1/m,1/(m-1)]$.
For $m\le 6$, this root can be obtained analytically, and for spectrally broader pulses, {\it i.e.} larger values of $m$ this root can be found numerically;
the solutions $\{\lambda, c^{opt}_j\}$ for $m\le 8$ are depicted in Table \ref{table}.

\begin{table}[!htb]
\hspace{6mm}\begin{tabular}{c||ccccccc}
$m$ & $2$ & $3$ & $4$ & $5$ & $6$ & $7$ & $8$\\ \hline
$\lambda$ & $\frac{2}{3}$ & $\frac{(6-\sqrt{3})}{11}\simeq 0.388$ &
$\frac{(5-\sqrt{5})}{10}\simeq 0.276$ &
$ 0.215$ &
$ 0.177$ &
$0.150$ & $0.130$\\ \hline
$c^{opt}_1$ & -0.144 &-0.032 & -0.015 & -0.009 & -0.006 & -0.004 & -0.003\\
$c^{opt}_2$ & 0.288 & -0.179 & -0.051 & -0.026 &  -0.016 & -0.011 & -0.008\\
$c^{opt}_3$ & 0 & 0.368 & -0.201 & -0.064 & -0.034 & -0.022 & -0.015\\
$c^{opt}_4$ & 0 & 0 & 0.435 & -0.218 & -0.073 & -0.040 & -0.026\\
$c^{opt}_5$ & 0 & 0 & 0 & 0.493 & -0.231 & -0.081 & -0.046\\
$c^{opt}_6$ & 0 & 0 & 0 &0 & 0.546 & -0.242 & -0.088\\
$c^{opt}_7$ & 0 & 0 & 0 &0 &0 & 0.595 & -0.252\\
$c^{opt}_8$ & 0 & 0 & 0 &0 &0 &0 & 0.641
\end{tabular}
\caption{Numerical values of optimal amplitudes $c^{opt}_j$ and $\lambda$ defined in Eq.(\ref{eq:optimalpulse}) and  Eq.(\ref{eq:lambda}) respectively for up to $m=8$ frequencies. Exact values of $\lambda$ for $m=5$ is $\lambda =\frac{1}{548}(3(75+\sqrt{145})-\sqrt{10830+802\sqrt{145}})$ and for $m=6$ is $\lambda=\frac{1}{252}(98+7\sqrt{7}-\sqrt{2891+868\sqrt{7}})$. }
\label{table}
\end{table}

\begin{figure}[!htb]
\hspace{40 mm}\includegraphics[scale=0.22]{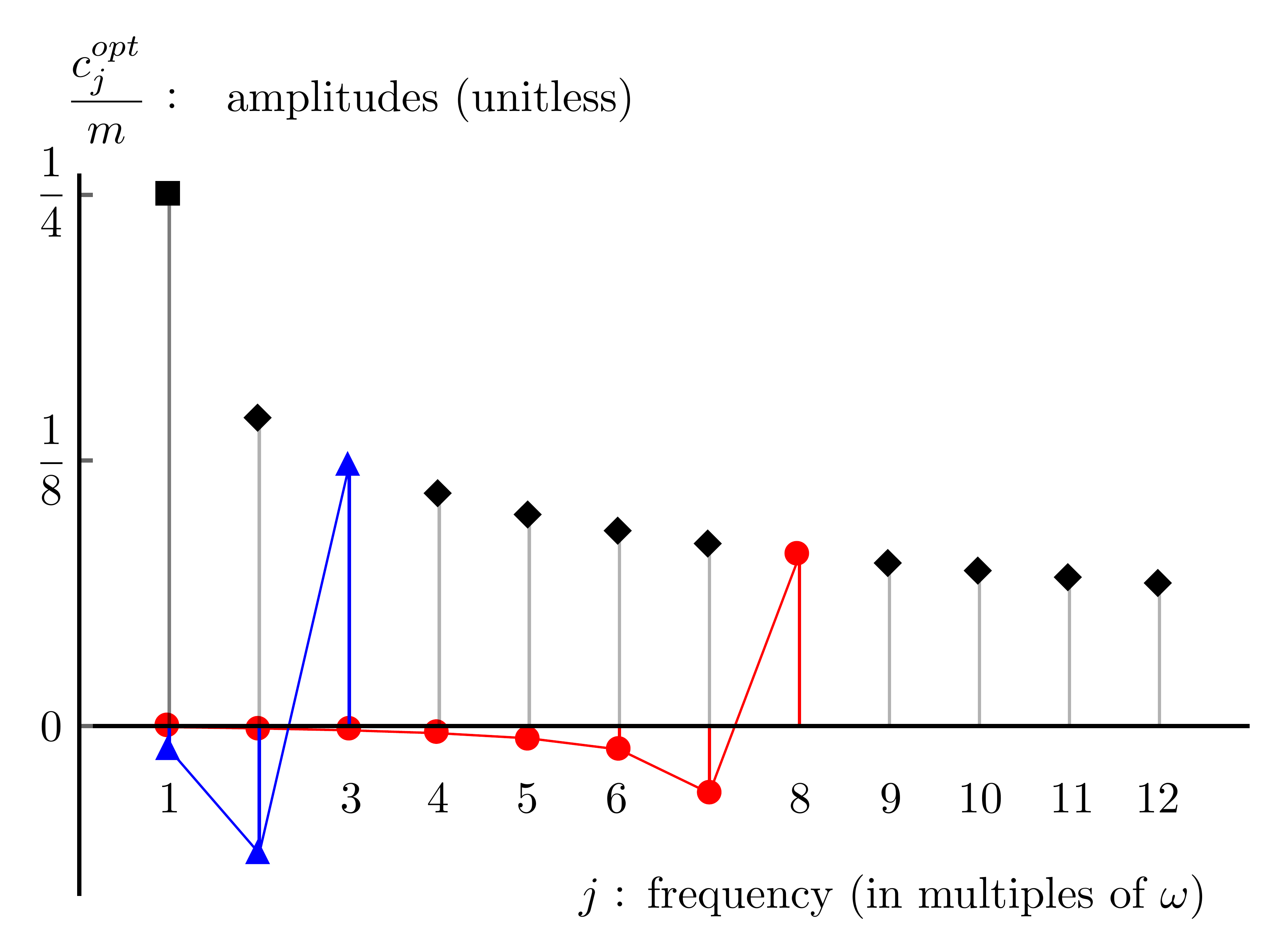}
\caption{(color online) The blue triangles (red circles) depict the amplitudes $c_j^{opt}/m$ of the optimal pulse (Eq.(\ref{eq:optimalpulse})) with $m=3\ (8)$ frequency components -- the line connecting the symbols is only to guide the eye.
The amplitude of the highest frequency component is always dominant and phase shifted as compared to all other amplitudes.
The filled square depicts the amplitude  for $m=1$ ({\it i.e.} a conventional MS gate) and the diamonds depict the amplitude of the highest (and dominant) frequency component for $c_m^{opt}/m$ for $m$ ranging from $2$ to $12$.
The conservative comparison $m\omega=$const. implies that gate duration increases with $m$ and the amplitudes $c_j^{opt}/m$ get lower with increasing $m$ as depicted here.}
\label{figure1}
\end{figure}

The exemplary cases for optimised pulses with $m=3$ and $m=8$ are depicted in Fig.(\ref{figure1}); independently of $m$, there is the general pattern that the absolute values of the amplitudes grow with increasing frequency and that there is a phase shift of $\pi$ between the amplitude of the highest frequency and all other amplitudes. 

With this solution, the most off-resonant process yields the dominant contribution to the gate performance,
and the phase shift of $\pi$ between the dominant amplitude and all other amplitudes results in destructive interference that suppresses undesired qubit-phonon entanglement
 as characterized by $f(t)$ in Eq.(\ref{eq:UMS}).

\subsection{Comparison with  conventional driving}
%monochromatic driving}

When comparing
 conventional and polychromatic driving one needs to bear in mind that the performance can always be improved through an increase of the fundamental frequency $\omega$,
 what, in practice however would require more intense lasers and would result in increased off-resonant excitations of additional transitions that decrease the gate fidelity.
In the following comparison we will thus always consider the case that the detuning $\delta$  in the conventional gate equals the highest frequency $m\omega$ of the polychromatic
pulse, so that the polychromatic
pulse contains no higher frequencies than  with conventional driving,
and the period of the polychromatic pulse is a factor of $m$ longer than that of the monochromatic pulse.

Since the impact of motional heating on gate fidelities  can always be reduced by increasing the gate duration \cite{PhysRevA.62.022311}, we will compare in the following gates of equal duration.
That is, in the conventional case, we consider gates with a duration of $m$ driving periods, whereas polychromatic gates are considered to be performed within one period of driving.

With this comparison, the intensity of the conventional driving ${\cal E}_{con}$ in terms of the fundamental driving frequency $\omega$ reads
${\cal E}_{con}=(\eta\Omega)^2=m\omega^2/16$ as compared to the intensity ${\cal E}_{pol}=\sum_{j=1}^m(c_j^{opt}\omega)^2$ with $c_j^{opt}$ given in Eq.(\ref{eq:optimalpulse}) for the optimised polychromatic case.
These intensities are of comparable magnitude, but one can see that ${\cal E}_{pol}$ never exceeds ${\cal E}_{con}$:
the exact value of ${\cal E}_{pol}$ is determined by the smallest root of Eq.(\ref{eq:lambda}),
which lies in the interval $[1/m,1/(m-1)]$.
In this interval, the function $\sum_{j=1}^m(c_j^{opt}\omega)^2$ decreases monotonically with $\lambda$; that is, one obtains an upper bound to the intensity by replacing the actual value of $\lambda$ by $1/m$, which (using Eq.(\ref{eq:optimalpulse})) results in
\be
{\cal E}_{pol}\le
\lim_{\lambda\to\frac{1}{m}}\frac{\omega^2\sum_{j=1}^m\frac{j^2}{(1-j\lambda)^2}}{16\sum_{j=1}^m\frac{j}{(1-j\lambda)^2}}=
\lim_{\lambda\to\frac{1}{m}}\frac{\omega^2\frac{m^2}{(1-m\lambda)^2}}{16\frac{m}{(1-m\lambda)^2}}=\frac{m\omega^2}{16}={\cal E}_{con}\ .
\ee
That is, in the following comparison, the polychromatic pulse will always be weaker and contain no higher frequencies than the monochromatic pulse.

\subsection{Displacement in phase-space}
The entangling gate operation is based on a displacement of the bus mode in phase space conditioned on the state of the qubit degrees of freedom.
This displacement is characterized by the real and imaginary part of the function $f(t)$ defined in the context of Eq.(\ref{eq:HMS}),
as it can be seen with the slight reformulation 
\begin{equation}
\mathcal{U}_{\mathcal{K}}=\exp\left(-i\big((\Re [ f(t)]x+\Im [f(t)]p)\mathcal{S}_x-g(t)\mathcal{S}_x^2\big)\right)\ ,
\label{eq:propagator}
\end{equation}
where $x=a+a^\dagger$ and $p=i(a-a^\dagger)$ 
are the operator valued part of position and momentum of the bus mode.

\begin{figure}[!htb]
\hspace{5mm}\includegraphics[scale=0.22]{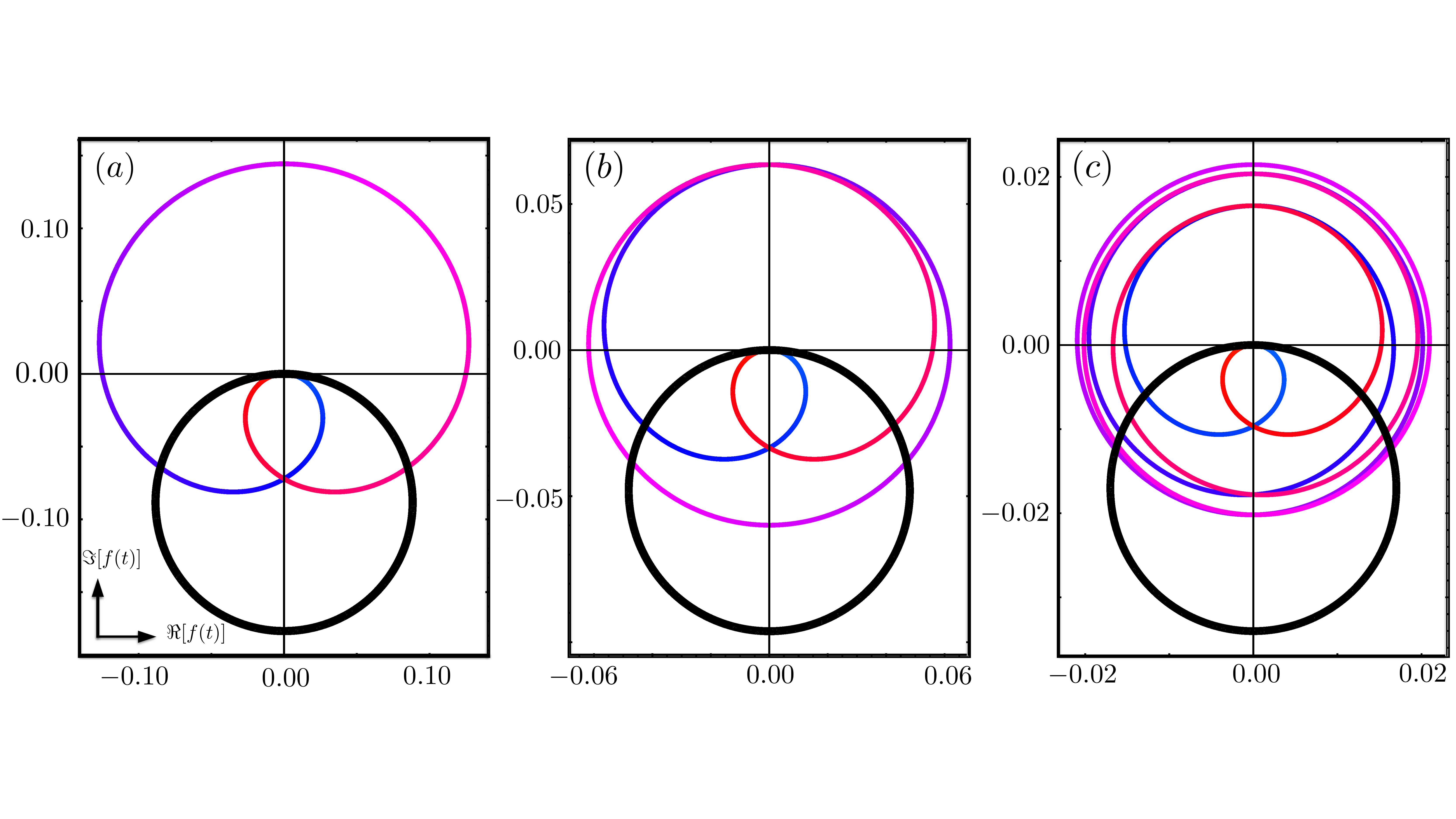}
\caption{
Phase-space trajectories of the bus mode characterized by $\Re [f(t)]$ and imaginary part $\Im [f(t)]$ (defined in Eq.\ref{eq:propagator}) for optimized gates with $m$ frequencies with $ m=2, 3, 6$, shown in $ (a), (b), (c)$ respectively (depicted in color changing from blue to red).
The corresponding trajectories for gates induced by monochromatic driving are depicted in black (the black trajectories are executed $m$ times). }
\label{figure}
\end{figure}
The real part $\Re [f(t)]$ and imaginary part $\Im [f(t)]$ are depicted in Fig.~(\ref{figure}) for optimized gates with $m=2,3$ and $6$ frequencies. The color ranging from blue to red indicates evolving time (blue colour corresponds to the beginning of the gate and red corresponds to the end of the gate).
The phase space trajectories for the corresponding monochromatic cases, as depicted in black, describe circular motion around a phase space point of finite displacement.
In the polychromatic case on the other hand, the trajectories are centred around the origin and the averaged displacement vanishes as enforced by the constraint $\av{f}$.
With increasing $m$ the trajectories are approximated better and better by circles around the origin, Since this results in overall reduced displacement, and the displacement couples the qubit degrees of freedom to the motional decoherence (following Eq.(\ref{eq:linbladoperators})),
the polychromatic driving results in increased gate fidelities as discussed more quantitatively in the following.

\subsection{Gate fidelities}

\begin{figure}[!htb]
\hspace{40mm}\includegraphics[scale=0.43]{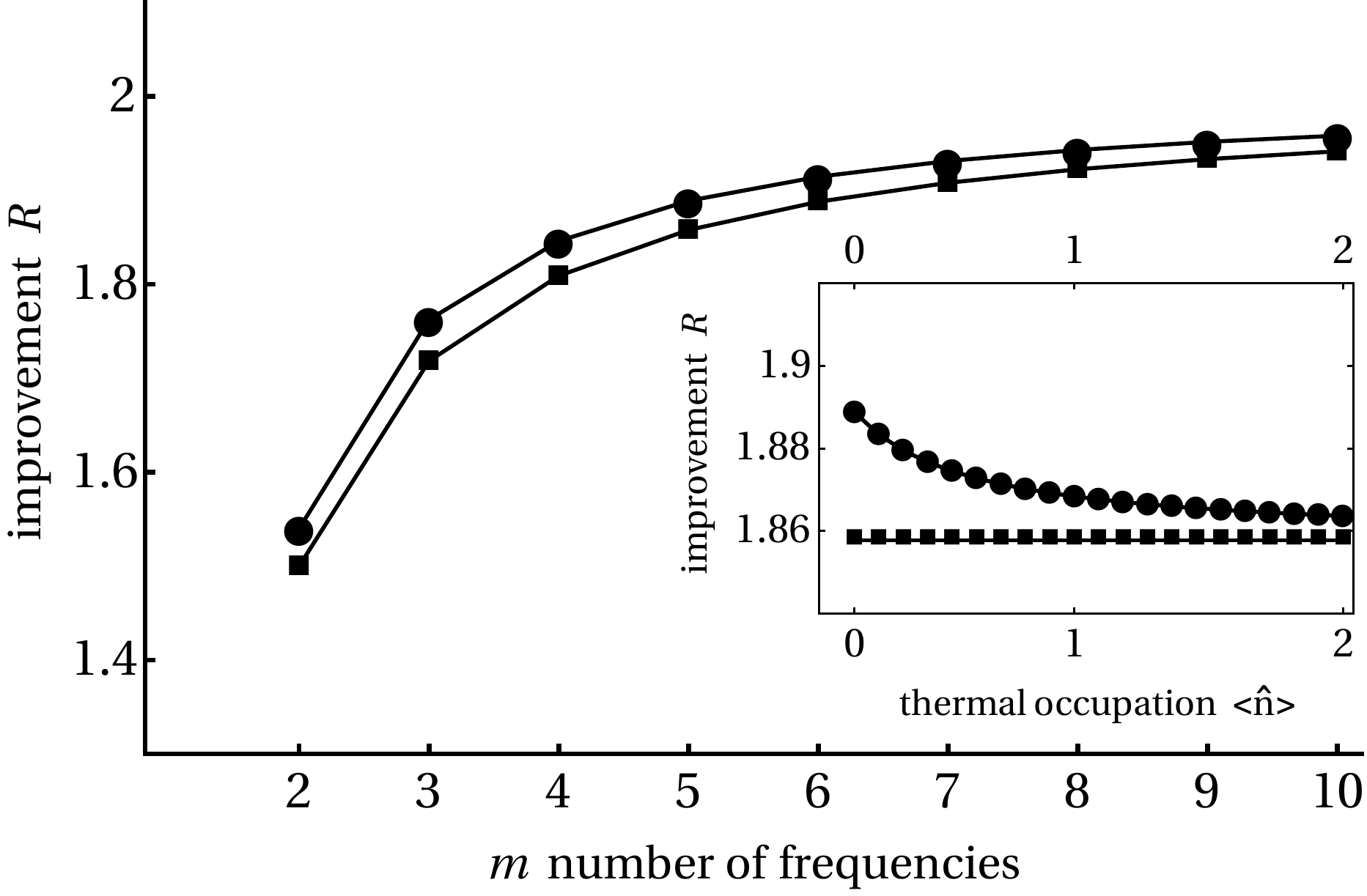}
\caption{Improvement $R=\mathcal{I}_m/\mathcal{I}_p$ of gate infidelities ($\mathcal{I}_m$ and $\mathcal{I}_p$ for monochromatic and polychromatic driving) in terms of  number of the frequencies in an optimized polychromatic pulse for a two-qubit gate, $N=2$, with a bus mode initially in the ground state.
Circles correspond to vanishing thermalisation and finite dephasing $\gamma_+=\gamma_-=0$, $\gamma_d>0$,
squares correspond to finite thermalisation  with equal rates and vanishing dephasing, $\gamma_+=\gamma_->0, \gamma_d=0$. The inset depicts the dependence of the improvement $R$ on the excitation $\langle\hat n\rangle$ for a thermal state of the bus mode and a pulse of $m=5$ frequencies.}
\label{figure2}
\end{figure}

Fig.(\ref{figure2}) depicts the improvement $R$ defined in terms of the ratio of the gate infidelities ${\cal I}$ for an optimized pulse and a monochromatic pulse, $\mathcal{I}_m/\mathcal{I}_p$, in a two-qubit system ($N=2$).
Circles and squares correspond to different environment models
($\gamma_+=\gamma_-=0$, $\gamma_d>0$ and
$\gamma_+=\gamma_->0$, $\gamma_d=0$,
with rates $\gamma_+$, $\gamma_-$ and $\gamma_d$ defined in Eq.(\ref{eq:lindblad})), and the improvement is plotted as a function of the number of frequencies $m$ in the pulse $\Upsilon_p$.
The bus mode is assumed to be in its ground state.
As one can see, a pulse with two frequencies (that permits to realize the condition $\av{f}=0$) yields an improvement by a factor around $1.5$  depending on the bus mode dissipation.
Increasing the number of frequencies enhances this improvement and values of $R$ around $2$ can be reached. 
 This factor of $2$ can intuitively be understood from Fig.~\ref{figure}: as $m$ tends to $\infty$ the phase space trajectories in the optimised case are perfect circles centred around the origin (for finite $m$ the circles are not perfect, but their radius changes in time),
and the radii for conventional and polychromatically driven gate coincide.
Since the conventional gate is based on circular trajectories around a point of finite displacement,
the optimised solutions permit to cover the same enclosed phase space area (formally expressed as $\sum_{j=1}^m|c_j|^2/j$) as required to realise the desired gate with a lower overall displacement.
In the formal framework, this shift of center is ensured via the constraint $\sum_{j=1}^m c_j/j=0$,
which directly enters the quantity
$\sum_{j=1}^m|c_j|^2/j^2+|\sum_{j=1}^mc_j/j|^2$ to be minimised
(as defined in Eq.~\ref{eq:g1} in the Appendix).
For optimised gates the second term vanishes, and the first term reduces to $\lambda/16$  defined via Eq.~\ref{eq:lambda}, and as $m\rightarrow \infty$,  $\lambda$ tends to $1/m$, and eventually the optimised quantity becomes $1/(16m)$; for the conventional gate however, the second term in the summation does not vanish and for the Rabi frequency $\eta\Omega=1/(4\sqrt{m})$ the whole summation becomes $1/(8m)$ which is twice the magnitude for the polychromatic gate.   

 In the case of initial phonon excitations, the gate fidelity is not merely a function of $\av{f}$ and $\av{f^2}$,
but also depends on higher moments of $f$, what results in a slightly different dependence on $m$,
as depicted in the inset of Fig.~(\ref{figure2}) that displays the dependence of $R$ on the mean phonon excitation $\langle\hat n\rangle$.
As one can see in this exemplary case of $m=5$, there is a slight drop of improvement with increasing $\langle\hat n\rangle$, but even in the limit of large excitations, improvements around $1.9$ are reached. For the environment model of $\gamma_-=\gamma_+=\gamma_d>0$ the obtained improvement (not shown here) lies in between the two curves depicted in Fig.(\ref{figure2}).

\begin{figure}[!htb]
\hspace{40mm}\includegraphics[scale=0.46]{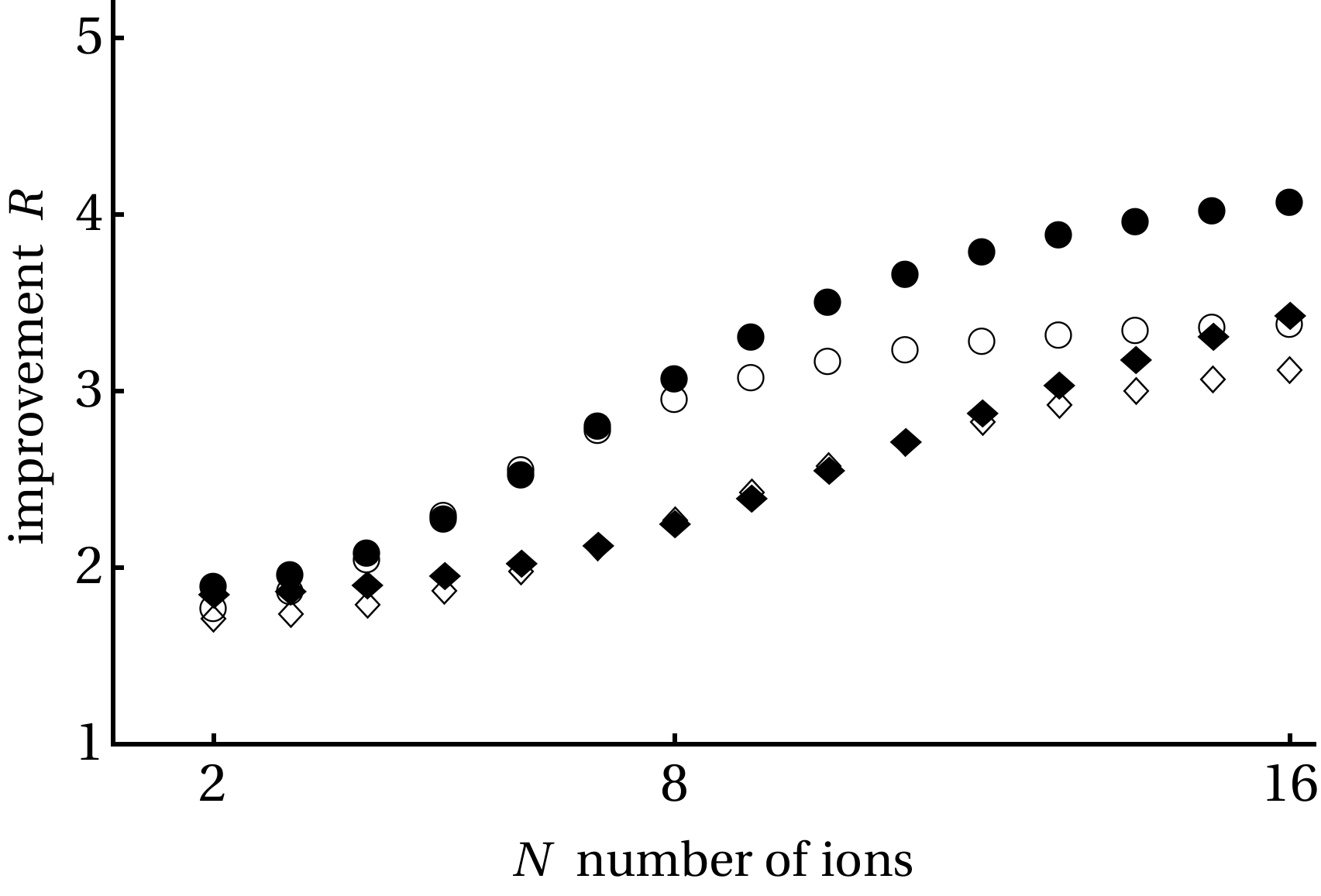}
\caption{Improvement $R$ as function of the number of ions for a pulse with $m=3(5)$ frequencies depicted by empty(full) symbols.
(circles correspond to $\gamma_+=\gamma_-=0$, $\gamma_d>0$;
diamonds correspond to $\gamma_+=\gamma_-=\gamma_d>0$)
The improvement rapidly grows with $N$ and the asymptotic improvement (for $N\to\infty$) grows with $m$.}
\label{figure3}
\end{figure}

As Fig.~\ref{figure3} shows, this improvement also grows  with the number of ions.
For a pulse with only $3$ frequencies $R$ reaches values  exceeding $3$,
and a spectrally broader pulse with $m=5$ yields values exceeding $4$. The dependence of $R$ on the number of ions $N$ can be understood in terms of the elementary processes that contribute to a reduction of gate fidelities, and those are characterized in terms of the matrix $\zeta$ (defined in context of Eq.(\ref{eq:map})) whose norm defines the gate infidelity.
According to the Eq.(15) in the appendix, 
the rates for the occurrence of simultaneous phase-flip 
induced by dephasing of the bus mode scale with $N^{4}$,  where $N$ is the number of ions.
Since most of these processes disappear for optimized pulses due to the constraint $\av{f}=0$, and the remaining contributions are being minimized, the improvement indeed grows with  increasing system size $N$.
Only for an environment model that does not contain any pure phase loss ({\it i.e.} $\gamma_d=0$), the improvement $R$ is indeed independent of the number of ions.

\subsection{Entanglement}

A central task of quantum gates is the creation of highly entangled states, but given the highly non-linear dependence of entanglement on the underlying quantum state, the gate fidelity itself does not necessarily provide an accurate assessment of the achievable entanglement. Therefore, we characterize the achievable entanglement of a two-qubit gate in terms of entanglement of formation (EoF) \cite{PhysRevLett.80.2245}. Since this is done with a numerical exact simulation that is not limited to the regime of weak dissipation, it particularly permits to assess the gate performance in strongly dissipative systems.
\begin{figure}[!htb]
\hspace{40mm}\includegraphics[scale=0.42]{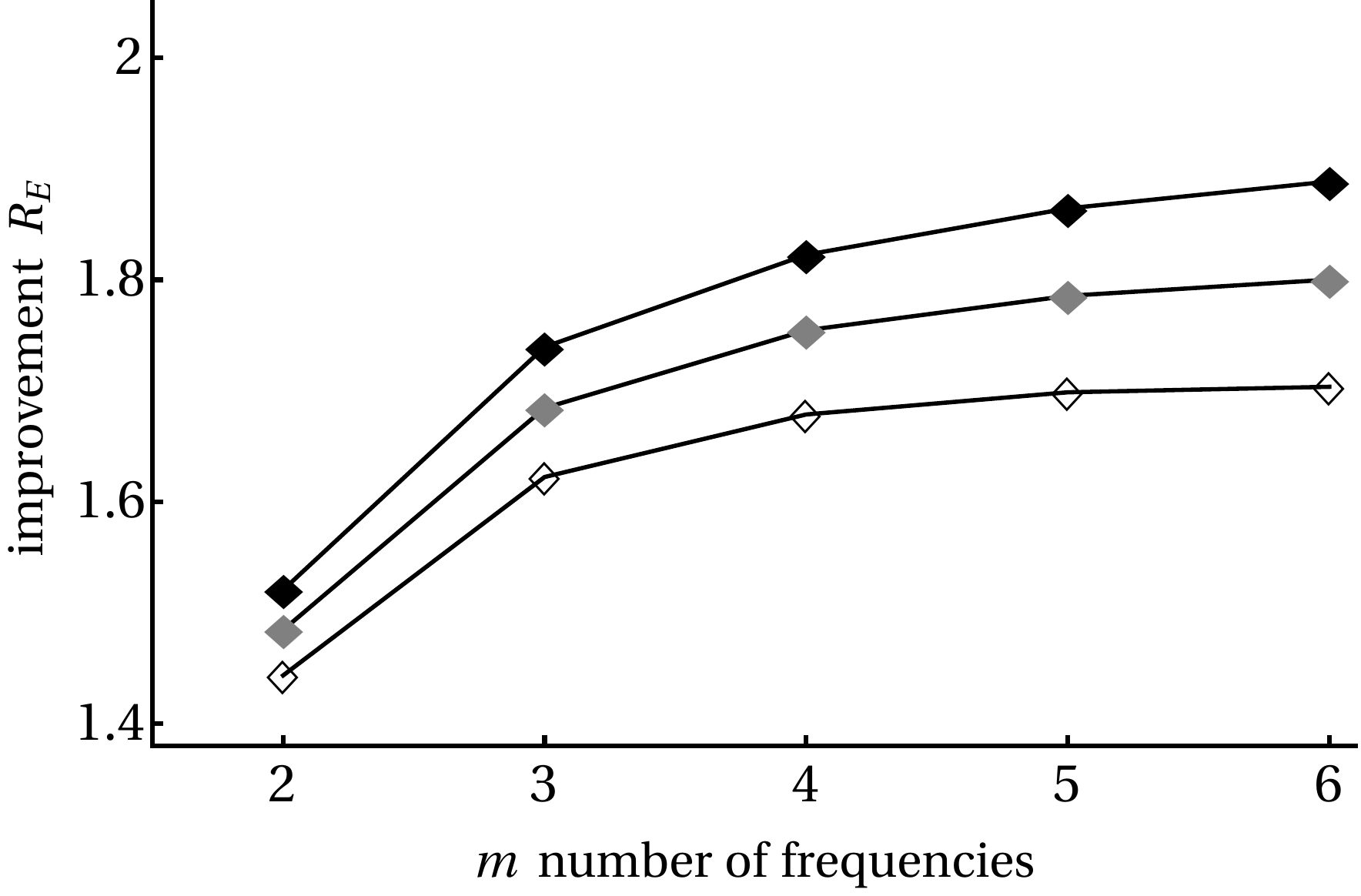}
\caption{Improvement $R_E$ of the deviation of EoF from its ideal value of $1$ in terms of the number of frequencies of an optimal polychromatic pulse for a two-qubit gate, $N=2$.
All data correspond to $\gamma_-=\gamma_+=\gamma_d$, with $\gamma_-=10^{-3}\delta$ depicted in solid, $\gamma_-= 5 \times 10^{-3}\delta$ depicted in grey  and, $\gamma_-=10^{-2}\delta$ depicted in empty diamonds.}
\label{figure4}
\end{figure}
Fig.(\ref{figure4}) depicts  the ratio $R_E=(1-E_m)/(1-E_p)$ of the deviation of EoF from 1 for a monochromatic and an optimized polychromatic two-qubit gate, where $E_m$ and $E_p$ denote EoF for the monochromatic and polychromatic gate respectively.
One can see that similarly to Fig.(\ref{figure2}), already pulses with $m=2$ frequencies yield a substantial improvement, and the improvement increases to a factor of $2$ for pulse with $m=6$ frequencies in the regime of weak dissipation (depicted by solid diamonds).
Quite importantly, even for stronger dissipation (depicted by empty and grey diamonds), there is still a substantial improvement.

\section{Conclusions}
Since the fidelities of entangling gates are the current bottleneck in the effort to reach fault tolerance,
the increased robustness with respect to dissipation in the bus mode
promises to advance control over trapped ions substantially. 
As shown here, it is possible to increase gate fidelities substantially, even in the presence of finite noise levels,
This fits the demands of modern surface trap architectures that bring ions close to noisy trap electrodes \cite{labaziewicz:180602,PhysRevLett.100.013001,ospelkaus2011microwave, PhysRevA.84.023412}, such that noise in the bus mode is a big challenge that is hard to overcome.

Given that the situation of an interaction between highly coherent systems that is mediated via an intermediate system with inferior coherence properties is by no means limited to trapped ions,
but is common in most hybrid architectures of designed quantum systems, the present control scheme promises to become a versatile and frequently employed tool of quantum control. Moreover, since in many systems rates for energy-conserving dephasing processes are substantially larger than processes in which energy between system and environment is being exchanged, one would thus expect to benefit from  significant growth of $R$ in most realistic situations. 

The observed increase of potential for control of gates involving many ions also offers an interesting angle on the deconstruction of quantum algorithm into individual gates.
This is typically done in terms of single-qubit and two-qubit gates, but three-qubit gates such as the Toffoli gate or gates involving more ions are rarely considered.
In situations in which $N$-qubit gates are realizable (such as trapped ions) and help to reduce the required number of elementary gates, there is also the prospect to benefit from the increase with qubit number in the  improvement of gate fidelities.

\section{Acknowledgments}
We are indebted to fruitful discussion with Winfried Hensinger, Matthias Keller, Klaus M\o{}lmer, Cord M\"uller, Joe Randall, Tobias Sch\"atz and Albert Verdeny Vilalta.
Financial support by the European Research Council within the project ODYCQUENT is gratefully acknowledged.

\section*{Appendix} 

In this appendix we provide some technical details on the gate infidelity in particular its dependence on the state of the bus mode, the driving pulse $\Upsilon_p$ and the number of ions.

The functions $\av{f}$, $\av{f^2}$, $\av{|f|^2f}$, $\av{|f|^2}$ and $\av{|f|^4}$ that characterize the gate fidelity, can be expressed as
\bqa
\av{f}&=&\frac{2\pi i}{\omega}g_1\ \nonumber,\hspace{1cm}\av{f^2}=-\frac{2\pi}{\omega}g_1^2\ \nonumber,\\
\av{|f|^2f}&=&-i\frac{2\pi}{\omega}(g_2-2f_1g_1-|g_1|^2g_1)\ \nonumber,\\
\av{|f|^2}&=&\frac{2\pi}{\omega}(f_1 + |g_1|^2)\label{eq:g1}\ ,\\
\av{|f|^4}&=&\frac{2\pi}{w}(f_2 - 4\Re g_2g_1^\ast +  |g_1|^4 + 4f_1|g_1|^2)\ \nonumber.
\eqa

in terms of the functions
\be\ba{ll}
\displaystyle g_1=\sum_{i=1}^{m}\frac{c_i}{i}\ ,& \displaystyle g_2=\sum_{ijp=1}^{m}\frac{c_ic_jc_p^\ast}{ijp}\delta_{i+j,p}\ \nonumber,\\
\displaystyle f_1=\sum_{i=1}^{m}\frac{|c_i|^2}{i^2}\ , & \displaystyle f_2=\sum_{ijpq=1}^{m}\frac{c_ic_jc_p^\ast c_{q}^\ast}{ijpq}\delta_{i+j,p+q}\ ,
\ea\ee
whith the (unitless) amplitudes that characterize the pulse $\Upsilon_p=\sum_{j=1}^{m}c_j\omega\exp(ij\omega t)$.

The matrix $\zeta$ whose norm yields the gate infidelity reads
\be
\zeta=\frac{\pi}{\omega}(\zeta_0+\zeta_1)\ ,\mbox{ with }\ \zeta_1=\sum_{i=1}^{6}\zeta_{1i}\nonumber
\label{eq:infidelitymatrix}
\ee
and
\bqa
\zeta_0&=&2f_1(\gamma_++\gamma_-+(2\langle\hat n\rangle + 1)\gamma_d)M_1-4f_2\gamma_d\GN(N-1) M_2\ \nonumber\\
\zeta_{11}&=&2|g_1|^2(\gamma_++\gamma_-+(2\langle\hat n\rangle + 1)\gamma_d)M_1\ ,\nonumber\\
\zeta_{12}&=&16|g_1|^2f_1\gamma_d\GN M_3\ ,\nonumber\\
\zeta_{13}&=&4\gamma_d(|g_1|^4-4\Re(g_1g_2^\ast))\GN(N-1) M_2\ ,\\
\zeta_{14}&=&4\Re(g_1^2\langle a^2\rangle)\gamma_d M_4\ ,\nonumber\\
\zeta_{15}&=&8\Re(g_1|g_1|^2\langle a\rangle)\gamma_d M_5\ ,\nonumber\\
\zeta_{16}&=&4\Im(g_1\langle a\rangle)(\gamma_+-\gamma_-+\gamma_d)M_6\ .\nonumber
\eqa
\\

Since the map defined in Eq.~\ref{eq:map} can be described in terms of five operators (the identity and the $\zeta_i$) defined in Eq.~\ref{eq:xi}), the coefficient matrices are five-dimensional (only a treatment of dissipative effects beyond first order would require the use of more than five operators and correspondingly larger matrices $M_i$.)
Explicitly the $M_i$ read

\bqa
M_1&=&\left[
\ba{ccccc}
-\GN&0&-\sqrt{\Gamma_2}&0&0\\
0&2\Gamma_1 &0&0&0\\
-\sqrt{\Gamma_2}&0&0&0&0\\
0&0&0&0&0\\
0&0&0&0&0
\ea
\right]\ ,\nonumber\\
M_2&=&\left[
\ba{ccccc}
1&0&\sqrt{\Gamma_2}&0&-3\sqrt{\Gamma_4}\\
0&0&0&0&0\\
\sqrt{\Gamma_2}&0&-2&0&0\\
0&0&0&0&0\\
-3\sqrt{\Gamma_4}&0&0&0&0
\ea
\right]\ ,\nonumber\\
M_3&=&\left[
\ba{ccccc}
-(N-1)&0&-\sqrt{\Gamma_2}(N-1)&0&3\sqrt{\Gamma_4}\\
0&0&0&0&0\\
-\sqrt{\Gamma_2}(N-1)&0&2(N-1)&0&0\\
0&0&0&0&0\\
3\sqrt{\Gamma_4}&0&0&0&0
\ea
\right]\ ,\nonumber\\
M_4&=&
\left[
\ba{ccccc}
0&0&\sqrt{\Gamma_2}&0&0\\
0&\GN&0&0&0\\
\sqrt{\Gamma_2}&0&0&0&0\\
0&0&0&0&0\\
0&0&0&0&0
\ea\right]\ ,\nonumber\\
M_5&=&
\left[
\ba{ccccc}
0&\sqrt{N}(\GN-1)&0&3\sqrt{\Gamma_3}&0\\
\sqrt{N}(\GN-1)&0&-\sqrt{\Gamma_1\Gamma_2}&0&0\\
0&-\sqrt{\Gamma_1\Gamma_2}&0&0&0\\
3\sqrt{\Gamma_3}&0&0&0&0\\
0&0&0&0&0
\ea\right]\ ,\nonumber\\
M_6&=&
\left[
\ba{ccccc}
0&-i\sqrt{N}&0&0&0\\
i\sqrt{N}&0&0&0&0\\
0&0&0&0&0\\
0&0&0&0&0\\
0&0&0&0&0
\ea\right]\ \nonumber.
\eqa

with $\Gamma_p= {N\choose p}$. Since $\zeta_{1}$ vanishes if $g_1$ (or, equivalently $\av{f}$) vanishes,
a large improvement in gate fidelity is achieved even with an only bi-chromatic pulse that permits to satisfy this constraint. 
$\zeta_0$ can be reduced only through minimization of $f_1$ and $f_2$, and thus will always be finite for a non-vanishing pulse.

\section{References}
\bibliography{ref.bib}

\end{document}